\documentclass[a4paper,11pt]{article}
\usepackage{jinstpub} 
\usepackage{lineno}

\usepackage{siunitx}
\DeclareSIUnit{\belmilliwatt}{Bm}
\DeclareSIUnit{\belfullscale}{BFS}
\DeclareSIUnit{\belcarrier}{Bc}

\usepackage[noabbrev]{cleveref}
\crefname{equation}{eq.}{eqs.}
\usepackage{subcaption}
\usepackage{svg}
\usepackage[extractname=filename]{svg-extract}

\title{\boldmath Application of RFSoC Technology for Beam Position Monitors at the SuperKEKB Storage Rings Injection Points}

\author[a,b,1]{B. Urbschat,\note{Corresponding author.}}
\author[b,c]{G. Mitsuka,}
\author[d]{L. Ruckman}
\affiliation[a]{Nagoya University,\\Furo-Cho, Chikusa, Nagoya, Aichi, Japan}
\affiliation[b]{KEK,\\Oho, Tsukuba, Ibaraki 305-0801, Japan}
\affiliation[c]{SOKENDAI,\\Shonan Village, Hayama, Kanagawa 240-0193, Japan}
\affiliation[d]{SLAC National Accelerator Laboratory,\\2575 Sand Hill Road, M/S 96, Menlo Park, CA 94025, USA}

\emailAdd{urbsch@hepl.phys.nagoya-u.ac.jp}

\abstract{
    In order to achieve its ambitious luminosity target, the SuperKEKB collider
    must achieve and sustain high beam currents on the order of Ampere in its
    storage rings.
    This requires continuous top-up injection and operation with a two-bunch
    injection scheme, injecting two $\SI{96}{\nano\second}$ spaced bunches in a
    single injection cycle. An important input for tuning the injection beam is
    the position reading from a dedicated beam position monitor (BPM), located
    after the septum magnets, slightly upstream of where the injected and
    stored beams converge.
    Previously, the readout electronics used for these special BPMs were not
    capable of independent measurement of both bunches in the two-bunch
    injection mode and modification of the concerned devices and their firmware
    was not feasible.
    The opportunity was taken to develop a new readout device based on the
    AMD/Xilinx RF System on a Chip (RFSoC) platform with the goal of not only
    providing a sufficiently flexible and performant readout solution for the
    concerned BPMs, but also to evaluate and gain experience with the platform
    for beam monitor electronics applications.
    This paper is concerned with the details of this development as well as 
    evaluation and operation of the developed device.
}


\arxivnumber{2602.14870} 

\begin{document}
\maketitle
\flushbottom

\section{Motivation and Requirements}
The SuperKEKB collider~\cite{superkekb_tdr} is an electron-positron collider
located at the High Energy Research Organization (KEK), Tsukuba, Japan.
At the time of writing, SuperKEKB already holds the world record for highest
instantaneous luminosity at $\SI{5.1e34}{\per\centi\meter\squared\per\second}$.
However, in order to support the physics goals of the accompanying Belle II
experiment~\cite{b2_design_report}, luminosities much beyond this value are
demanded.

Operation at such high luminosities requires considerable beam currents on the
order of multiple Ampere.
This, combined with the rather short beam life on the order of tens of minutes
necessitates efficient, continuous (top-up) injection into both of the
collider's storage rings.
The injection frequency is however limited by the present infrastructure to
maximally $\SI{25}{\hertz}$. 

As a means to maximize injected charge given such limitations, a
\textit{two-bunch injection mode} was implemented~\cite{twobunch_injection}.
During two-bunch injection, two $\SI{96}{\nano\second}$ spaced bunches are
accelerated during the same cycle of the injector linac and injected into two
appropriately spaced bunches in either of the storage rings.
This scheme effectively doubles the injected charge of around
$\SI{2}{\nano\coulomb}$ for the electron beam and around
$\SI{3}{\nano\coulomb}$ for the positron beam to twice as much per injection
cycle.

To optimize injection efficiency (amount of stored charge compared to injected
charge) the parameters of the injection (septum and kicker magnets) must be tuned.
As an input to the tuning a special beam position monitor (BPM) located just
upstream of the injection point is used to monitor the trajectory of the
injected bunches.
Notably the trajectories of both bunches in two-bunch injection mode may differ
and must therefore be tuned separately. This of course necessitates separate
measurement of both trajectories while taking care to ensure independent
measurements, meaning no overlap of signals from the first and second bunch,
which would distort measurements.
The present commercial BPM readout electronics were found to provide
insufficient flexibility for this rather special application, leading us to
pursue the in-house developed solution presented here.

The in-house development is further motivated by a general interest in adoption
of the \textit{RF System on a Chip} (RFSoC) platform by
AMD/Xilinx~\cite{AMDXilinx:wp489} as a flexible solution for future beam
monitor electronics, especially considering the excellent performance of the
included analog-to-digital converters (ADCs) paired with the possibility for
rapid development thanks to the single-chip adaptive solution.

In the following, we first provide an overview of the used hardware spanning
the BPM vacuum chamber as well as the adopted method for position calculation,
used 3rd generation RFSoC evaluation board and analog frontend circuit.
As part of the introduction of the RFSoC evaluation board, a measurement of the
effective number of bits of the 3rd generation RFSoC's ADCs measured using that
board is presented.
Next, we provide a high level overview of the custom firmware and software
developed for our application.
The following section focuses on evaluation of the developed readout device and
its use during accelerator operations. We present results of an early
resolution estimation study conducted at the KEK injector linac during the
development phase, address necessary channel gain corrections as well as give a
summary of use during recent SuperKEKB operations.
We conclude with a short summary and prospects for future developments of the
application concerned by this paper as well as RFSoC based electronics for beam
monitors at KEK in general.

\section{Hardware}
\subsection{SuperKEKB injection point BPM vacuum chamber}
\label{sec:bpmchamber}
\begin{figure}[htbp]
\centering
    \begin{subfigure}[t]{.48\textwidth}
        \centering\includegraphics{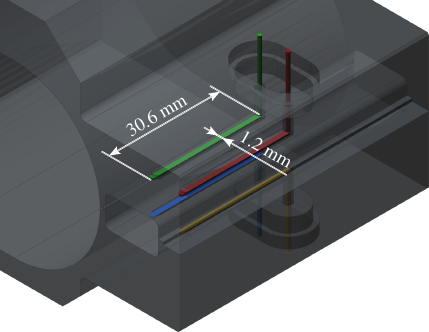}
        \caption{Oblique transparent view}
    \end{subfigure}
    \hfill
    \begin{subfigure}[t]{.35\textwidth}
        \centering\includegraphics[width=\linewidth]{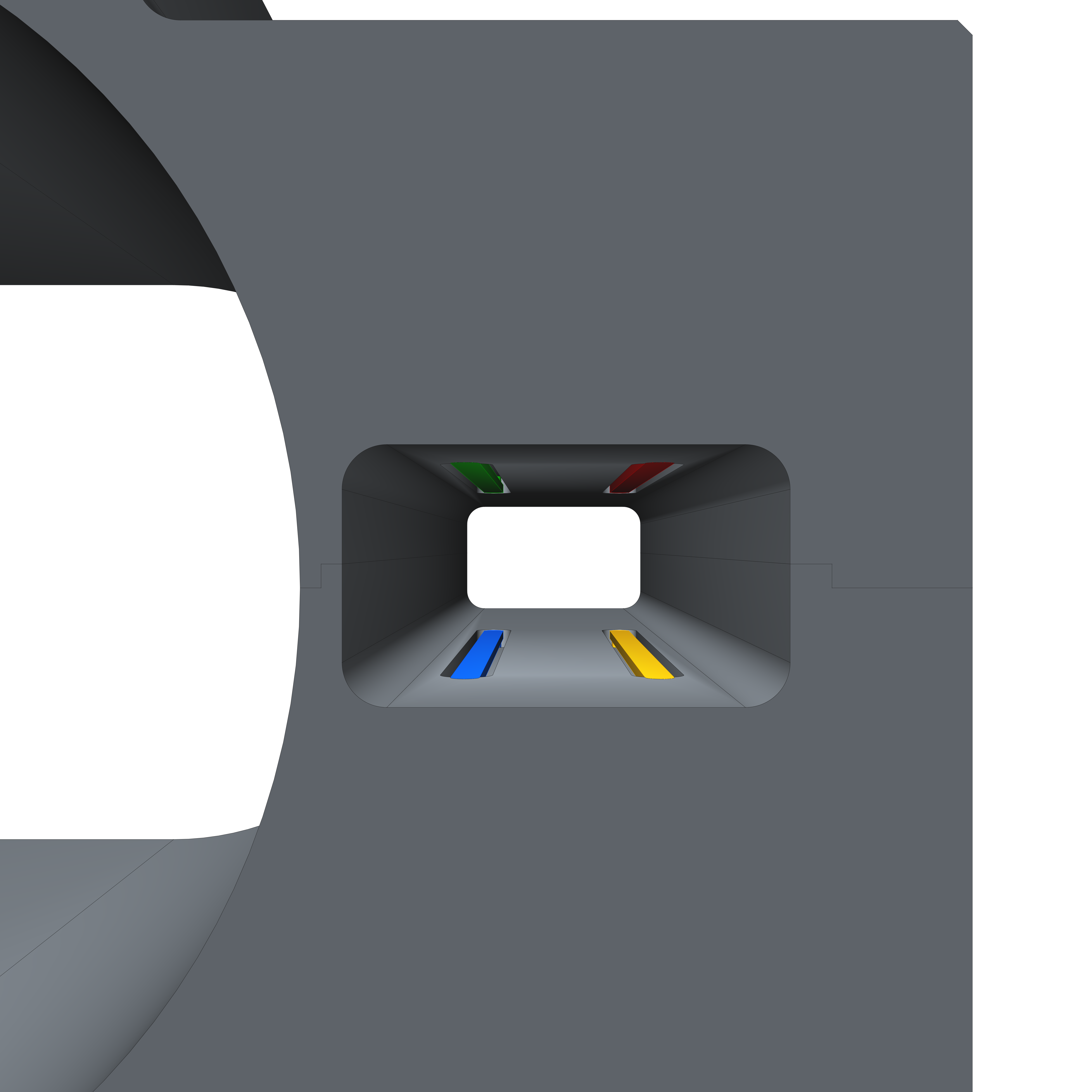}
        \caption{Head-on view}
    \end{subfigure}
\caption{Simplified 3D model of the injection point BPM vacuum chamber for the electron ring.\label{fig:bpmchamber}}
\end{figure}

Both the electron and the positron ring of SuperKEKB feature a special
\textit{injection point BPM}, measuring the trajectory of the injection beams
just upstream of the injection point where storage and injection beam vacuum
chambers converge and the injection beam meets the storage beam.
As shown in \cref{fig:bpmchamber}, the injection beam duct is attached to
the side of the storage beam's chamber as at the location of the BPM injection
and storage beam are already brought into close proximity.
Four pickup electrodes of stripline type, shorted to the chamber's body at one
end and connected to a feedthrough at the other, are placed at the top and
bottom of the injection beam duct.

The injection beam duct is rather small, sized
$\SI{8}{\milli\meter}\times\SI{15}{\milli\meter}$ for the electron ring and
$\SI{8}{\milli\meter}\times\SI{16.6}{\milli\meter}$ for the positron ring and
the beam does not necessarily traverse it near its center.
Given the rectangular geometry and rather wide range of possible
beam positions, it is not feasible to analytically derive a
relationship between the signals from the electrodes and beam
position.

Alternatively a map of signal strengths as a function of beam position may be
obtained in discrete form, either experimentally (by stretched wire scan) or
by adequate simulation. 
The discrete map can then be interpolated wherever a continuous one is
required. 
As no stretched wire scan of the chambers was performed before installation, we
are forced to rely on results from simulation. Specifically a custom simulation
code based on the Boundary Element Method (BEM) in two dimensions is used. Due
to the nature of this simulation only the signal strength in arbitrary units
but \textit{not} the shape of the signal waveform is determined.
The signal strength is assumed to be proportional to the integral of the
absolute values of the digitized waveforms. An absolute scale is not required
for position measurement. It would be a requirement for an absolute charge
measurement, however in our application a relative measure of bunch charge
suffices.

\begin{figure}[htbp]
\centering
    \begin{subfigure}[t]{.48\textwidth}
        \centering\includegraphics[width=\linewidth]{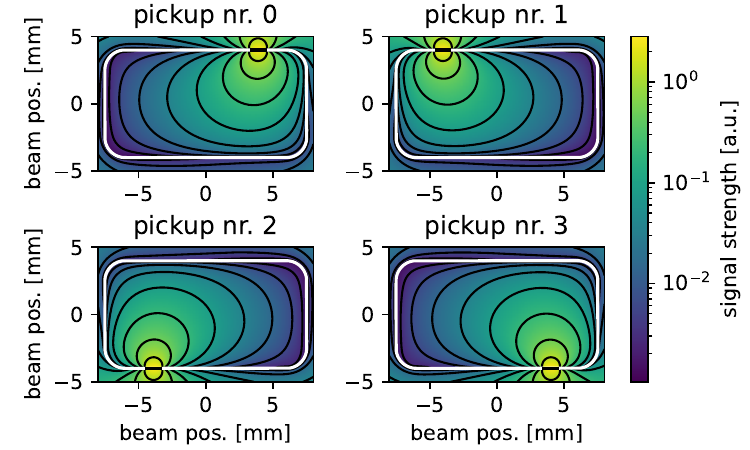}
        \caption{Electron ring}
    \end{subfigure}
    \quad
    \begin{subfigure}[t]{.48\textwidth}
        \centering\includegraphics[width=\linewidth]{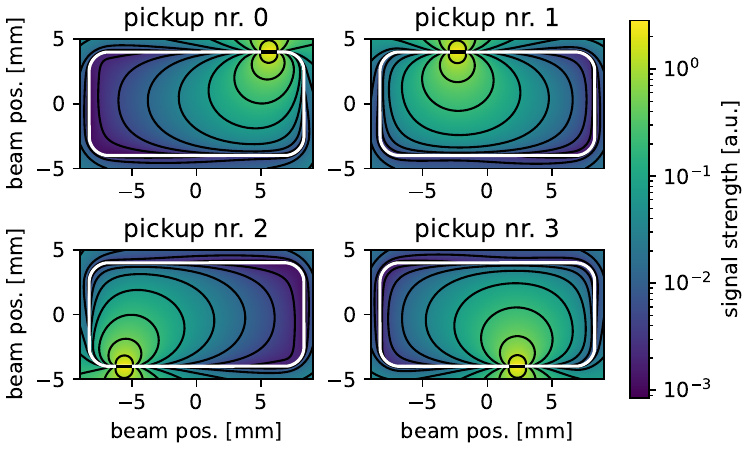}
        \caption{Positron ring}
    \end{subfigure}
\caption{
    Signal maps for all electrodes of the used BPM chambers as determined by BEM
    simulation. The used chamber outline is shown in white. Only the map values
    within the outline are of relevance.
    Black lines represent logarithmically spaced contours.
    \label{fig:bem_maps}
}
\end{figure}

The obtained signal maps with overlaid chamber cross-sections for both rings
are shown in \cref{fig:bem_maps}.
The electrodes are modeled as a section of the chamber wall. In reality, while
the electrodes do sit flush with the chamber wall, there is a small gap on
their sides and back to isolate them from the chamber wall.
Considering manufacturing tolerances and other factors, it was deemed
unnecessary to model fine details like the small gaps, as they are likely to
not exactly represent reality to begin with. This does however lead to
disagreements between simulated and true signal response, which become
significant for beam positions near the electrodes as further discussed in
\cref{sec:gain_calibration}.

Note that the chamber geometries slightly differ between electron and positron
ring. The positron ring's chamber is of an older type where the electrodes were
arranged asymmetrically in an attempt to improve sensitivity near the edges of
the chamber.
During recent operations this was not found to provide any advantage and rather
only introduce unnecessary complications. Thus, the electron ring's chamber was
already exchanged for a new one with symmetric electrode arrangement. As the
new chamber geometry was found to work well, the same will be installed for the
positron ring in the future.

The signal maps form the base for computation of the beam position as a
function of the four signal strengths from the four electrodes.
It was attempted to approximate this relationship using polynomials in four
variables (the signal strengths), one for each direction in the plane.
While this is expected to work well for beam positions near the center of the
duct, for the full range of expected beam positions in our application, even
with polynomials of very high degree (12 and above), only unsatisfactory
results could be achieved.

Thus, instead an iterative algorithm is adopted to find the position
coordinates $x, y$ for which the signal strengths determined by simulation are
closest to the measured ones.
The coordinates $x, y$ are determined such that the quantity $f(x, y)$ as
defined in \cref{eq:minimizefunc} is minimal. $s^{\text{map}}_i(x, y)$ is the
signal strength for the $i$-th electrode obtained for a given position $x, y$
from the interpolated signal strength map.
$s_i$ is the measured signal strength for the $i$-th electrode.
\begin{equation}\label{eq:minimizefunc}
    f(x, y) = \sum_{i = 0}^{3} \left( \frac{s^{\text{map}}_i(x, y)}{\Sigma_{i = 0}^{3} s^{\text{map}}_i(x, y)} - \frac{s_i }{\Sigma_{i = 0}^{3} s_i} \right)^2
.\end{equation}
For the minimization a C++ implementation of the \textit{BOBYQA}
algorithm~\cite{powell2009bobyqa, pitzl2018} is used. This algorithm runs
sufficiently fast to meet the $\SI{25}{\hertz}$ injection rate requirement with
ease.

Finally, it shall be noted that due to the placement of the used BPM chamber, a
meaningful in-situ assessment of the resolution by the 3-BPM~\cite{kekb_instr}
method is difficult, as one would have to reliably determine and reflect the
effect of the septum magnets located between the injection point BPM and other
BPMs located further upstream.

While not directly comparable, an estimation of resolution when using the
developed readout was carried out at the KEK injector linac, using the BPM
chambers installed there. Results from this study are presented in
\cref{sub:reseval_linac}.

\subsection{RFSoC 4x2 evaluation board}
\begin{figure}[htbp]
    \centering
    \includegraphics[width=0.5\textwidth]{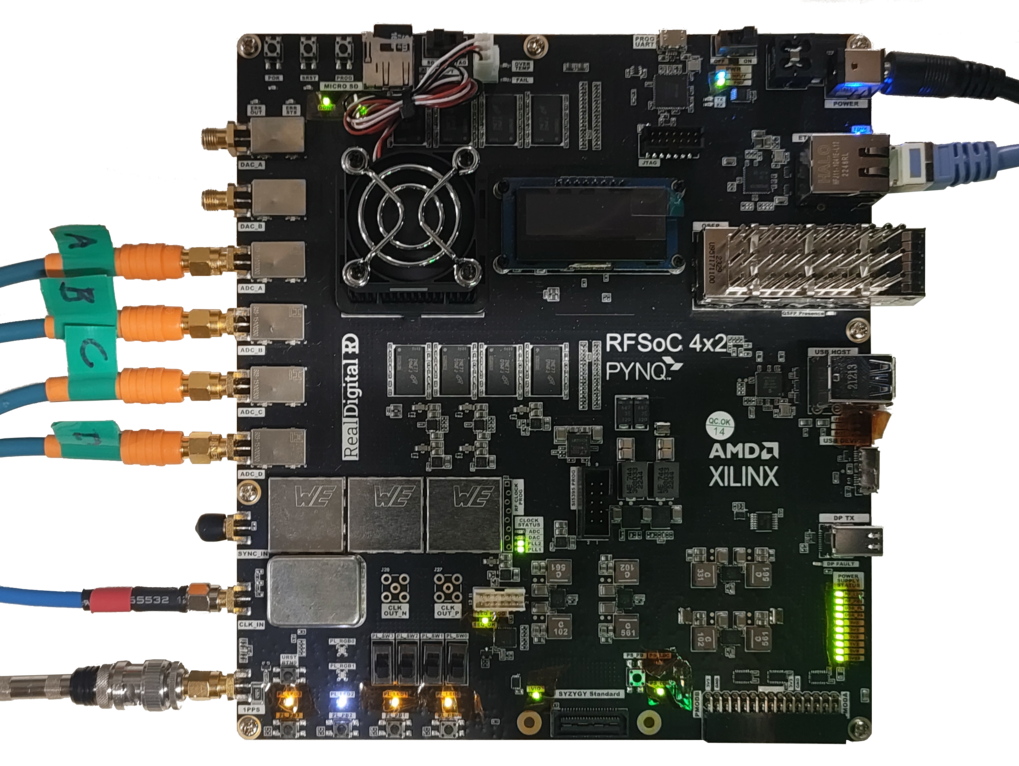}
    \caption{Photograph of the used RFSoC 4x2 evaluation board.
    \label{fig:rfsoc4x2_photo}}
\end{figure}

An \textit{RFSoC 4x2} evaluation board designed and distributed by RealDigital
is used to read out the signals from the BPM's pickups.
This board features a 3rd generation RFSoC containing high performance ADCs and
digital-to-analog converters (DACs), ARM processors as well as a field
programmable gate array (FPGA) fabric united in a single package.
The used evaluation board provides access to four of the integrated 14-bit, 5
gigasample per second (GSPS) ADCs and two of its 14-bit, 9.85 GSPS DACs.
Further inputs are available for trigger and clock signals.
A photograph of the board is shown in \cref{fig:rfsoc4x2_photo}.

\paragraph{Effective Number of Bits}
For our application the \textit{effective number of bits} (ENOB) of the ADCs, a
measure of the actual resolution of an ADC, is an important performance metric.
I. Degl’Innocenti et al. report an ENOB of around 9.5 in their survey of recent
ADCs~\cite{degl:ipac2023-thpl089}, which is extracted from the device's
datasheet.
We are not aware of any publications verifying the ADC performance by
measurement of the ENOB (or equivalent measurement) for a 3rd generation RFSoC.
The ADCs for the 3rd generation are 14-bit ADCs while previous generations had
12-bit ADCs. Especially for the 1st generation chips some detailed evaluations
have been published~\cite{10.1093/mnras/staa3895}.
Given the lack of an independent verification of the ADCs performance, we
decided to conduct an ENOB measurement using the RFSoC 4x2 board.
The ENOB is computed from the \textit{signal-to-noise and distortion ratio}
(SINAD) as described in~\cite{kester2009}.
SINAD captures the dynamic performance of an ADC as the ratio of a single
tone's power (signal power) to the power in the remaining spectrum, however
excluding DC (sum of noise and distortion power). We convert SINAD to ENOB as
that allows for a more intuitive interpretation.

An Agilent E8663D signal generator as well as bandpass filters had to be
employed to achieve sufficiently clean single tones.
Without the bandpass filters the measurement was strongly influenced by noise
and distortion from the signal generator, leading to measured ENOB of up to two
bits lower than what was measured with the filters.
Further, even with bandpass filters in place, some noise may leak through which
still can have an influence on the measurement. Initially an Agilent N5181B was
used which however had a high enough noise floor for frequencies below around
$\SI{300}{\mega\hertz}$ to significantly degrade the measured ENOB.

Bandpass filters were available for $\SI{150}{\mega\hertz}$ (Mini-Circuits
\textit{SBP-150+}), $\SI{1.5}{\giga\hertz}$ and $\SI{2}{\giga\hertz}$ (custom
coupled stripline).
Signal power is set so as to achieve the largest possible signal at the ADCs
which is still within their input range. For the bandpass filters at
$\SI{1.5}{\giga\hertz}$ and $\SI{2}{\giga\hertz}$ the insertion loss was very
large and even at the maximum output power of the signal generators of
$\SI{18}{\deci\belmilliwatt}$ only around a third of the ADCs input range could
be used. This however does not directly pose a problem as in the analysis the
power of each tone is measured relative to the full ADC scale and appropriate
scaling is applied.
The ADC sampling clock is locked to the $\SI{10}{\mega\hertz}$ reference signal
of the signal generator by configuring the dedicated phase-locked loops (PLLs)
on the RFSoC 4x2 board accordingly. The ADC sampling frequency is chosen as
$\SI{4.072}{\giga\hertz}$, an integer multiple of $\SI{509}{\mega\hertz}$,
which is close to the SuperKEKB global RF clock frequency the board will
eventually reference during operation. The number of samples in a buffer is
chosen as $8144$ , which divides the sampling rate. Measurements then may be
taken at the frequencies $f_n = n \cdot \SI{4.072}{\giga\hertz} / 8144$ where
$n = 1, 2, 3, \ldots$ and $f_n < \SI{2}{\giga\hertz}$.
This condition ensures a whole number of cycles of the sampled sinusoidal
signal in the buffer, thus eliminating the need for windowing functions for the
discrete Fourier transform. Windowing the signal has an effect on its power
spectrum, which would have to be corrected. By using a setup which does not
require windowing in the first place, such complications are circumvented.

In order to assess statistical fluctuations, the measurement is repeated 50
times at each frequency and for each channel.
The mean over the 50 measurements with one standard deviation as the
uncertainty for each channel is presented in \cref{tab:enob_meas}.

\begin{table}[htbp]
\centering
\caption{
    ENOB measurements for the 3rd generation RFSoC ADCs in each of the four
    channels (A, B, C, D) available on the RFSoC 4x2 board.
\label{tab:enob_meas}}
\smallskip
    \begin{tabular}{|c|c|c|c|c|c|}
        \hline
        Frequency & Filter & \multicolumn{4}{c|}{ENOB} \\
        \cline{3-6}
        & & A & B & C & D \\
        \hline
        $\SI{150}{\mega\hertz}$ & SBP-150                  & $10.08 \pm 0.02$ & $10.00 \pm 0.01$ & $10.06 \pm 0.01$ & $ 9.97 \pm 0.02$\\
        $\SI{1.5}{\giga\hertz}$ & coupled stripline & $10.16 \pm 0.01$ & $10.07 \pm 0.02$ & $10.11 \pm 0.01$ & $10.03 \pm 0.01$\\
        $\SI{2}{\giga\hertz}$   & coupled stripline & $10.05 \pm 0.02$ & $ 9.96 \pm 0.02$ & $10.04 \pm 0.02$ & $ 9.95 \pm 0.02$\\
        \hline
    \end{tabular}
\end{table}

Degl’Innocenti et. al.~\cite{degl:ipac2023-thpl089} use the performance metric
$P = 2^{\text{ENOB}} \cdot f_{\text{s}}$ to compare different ADCs based on
values extracted from corresponding datasheets. The input signal frequency is
chosen as close as possible to half of the maximum sampling rate
$f_{\text{s}}$. The best ADCs score a value of around
$P_{\text{best}}=\SI{2.90e12}{\per\second}$. Our measurement for an input
signal near $\SI{2}{\giga\hertz}$ implies around
$P\approx\SI{4.2e12}{\per\second}$ computed with $f_{\text{s}} =
\SI{4.072}{\giga\hertz}$ which is the sampling frequency for our setup.

The measured ENOB values and resulting performance metrics even slightly exceed
the values suggested by the datasheet. This confirms that the 3rd RFSoC ADCs
are indeed performing excellently and represent the state of the art of ADC
technology.

\paragraph{Digital Step Attenuators}
The 3rd generation RFSoCs, as used on the RFSoC 4x2 board, include
\textit{digital step attenuators} (DSAs) in each analog signal path providing
$\SI{0}{\deci\bel}$ to $\SI{27}{\deci\bel}$ attenuation adjustable in
$\SI{1}{\deci\bel}$ steps. Although not yet implemented, we explored the
possibility to employ the DSAs to dynamically set the attenuation for each
signal channel independently to optimize utilization of the ADCs dynamic range.
This is especially interesting in our use case where the beam is likely to
traverse the BPM chamber off-center, leading to a large asymmetry in the signal
amplitudes.
For this however it is vital to understand the characteristics of each
attenuator in order to correct for possible differences between the set and
true attenuation.

\begin{figure}[htbp]
    \centering
    \includegraphics[width=0.75\textwidth]{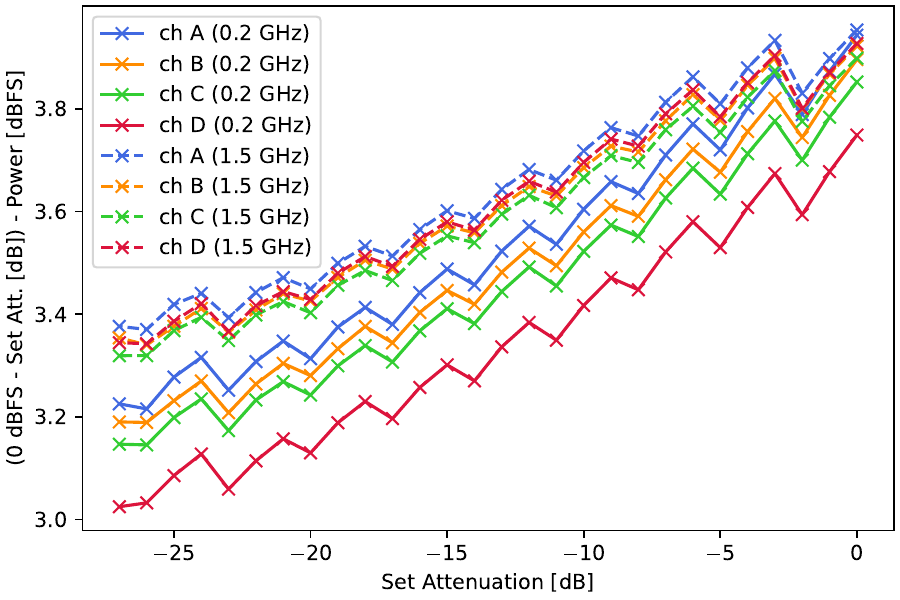}
    \caption{Difference between expected and measured signal power (up to a
        constant offset) as a function of the set attenuation of the step
        attenuator in each of the four channels available on an RFSoC 4x2
        evaluation board.
    \label{fig:step_attenuators}}
\end{figure}

A simple scan of attenuation values over the available range showed that the
set attenuations appear to be accurate to within about $\SI{1}{\deci\bel}$.
\Cref{fig:step_attenuators} shows the difference between measured and expected
power (up to a constant offset) as a function of the set attenuation in each
channel available on the RFSoC 4x2 board, measured at $\SI{200}{\mega\hertz}$
and $\SI{1.5}{\giga\hertz}$\footnote{Further frequencies were checked, but are
not shown in \cref{fig:step_attenuators} to avoid clutter.}.
A constant offset remains as in the calculation we reference
$\SI{0}{\deci\belfullscale}$ (decibels relative to full scale) but the input
signal did not actually fill the whole ADC range. It is the same for all four
channels as the signal power was not changed between measurements.
The measurements were performed with a sinusoidal input signal and the power is
extracted from the discrete Fourier transform of the recorded signal using the
same procedure (clock synchronization, buffer size) as used for the ENOB
measurement presented above. Power at the signal generator was set to
$\SI{4.80}{\deci\belmilliwatt}$.

Clearly there are some relative offsets on the order of tenths of a decibel
between the four channels. This may be attributed to slight differences in the
signal path, which could also be related to the signal traces on the printed
circuit board. Interestingly, up to an offset, at a given frequency the curves
in all four channels take almost exactly the same shape.
This indicates good relative accuracy between the attenuation elements used in
the step attenuator of a given channel.
A sawtooth-like pattern is observed, repeating every three decibels (i.e. every
three steps). This is likely related to a given attenuation element being
switched in or out of the signal path. The period of three rather than a power
of two further suggests that the step attenuator is likely to be a little more
sophisticated than a simple series of switched $2^{n} \si{\deci\bel}$ ($n = 0,
1, 2, \ldots$) attenuation elements, which also would not fit in with the
maximum attenuation of $\SI{27}{\deci\bel}$.

While a very similar pattern appears when measuring at different frequencies,
the relative offsets do exhibit frequency dependence. Further, the overall
slope towards lower attenuations appears ever so slighly shallower for the
higher frequencies. This means that while a frequency dependence exists, it is
for the most part independent of the attenuator setting.
In any case, this can in principle be corrected for. If further the sampled
signal does only change in amplitude but its spectral composition remains
constant, the corrections are much simpler, as in this case the corrections can
be applied directly in the time domain, while in general they may have to be
applied in the frequency domain.
In our application, the signal bandwith is well defined and narrow enough to
make corrections in only the time domain sufficient.

Further, given the observed relative offsets even at fixed frequency, it is
clear that in our application, if attenuation is set separately in each channel,
the attenuators must be properly characterized in order to apply attenuation
value dependent compensation. If however attenuators in all four channels are
set to the same value (and constant offsets between channels are accounted
for), an attenuation value dependent compensation may be safely omitted.

\subsection{Analog circuit and signal conditioning}
\begin{figure}[htbp]
\centering
    \begin{subfigure}[t]{.48\textwidth}
        \centering\includegraphics[width=\linewidth]{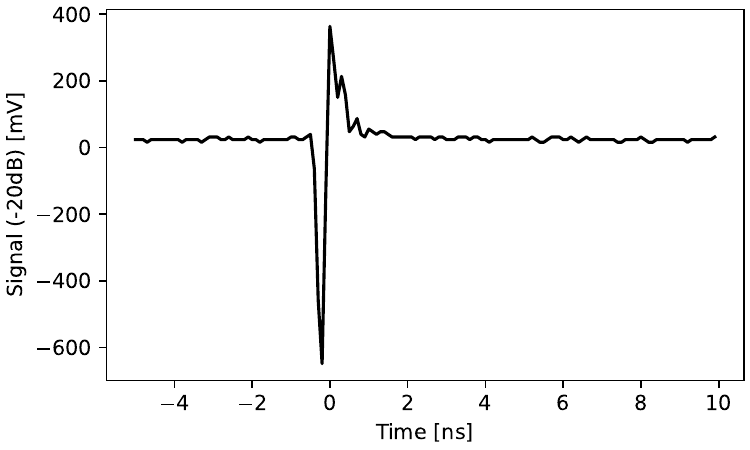}
        \caption{Raw signal (single bunch)}
    \end{subfigure}
    \quad
    \begin{subfigure}[t]{.48\textwidth}
        \centering\includegraphics[width=\linewidth]{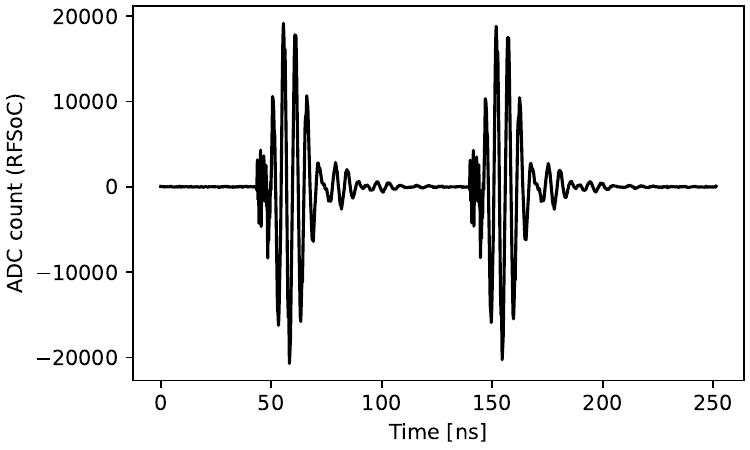}
        \caption{Shaped signal (two bunches)}
    \end{subfigure}
    \caption{Raw and shaped waveform from one of the electrodes. The former is attenuated by
        $\SI{20}{\deci\bel}$ and captured using a Rohde\&Schwarz \textit{RTO 1024}, 10 GSPS
        oscilloscope. The latter is captured using the RFSoC 4x2 board. Note
        the large difference in time scales.
    \label{fig:signal_conditioning_raw}}
\end{figure}

The signal as obtained from the pickup electrodes is a rather short
($\sim\SI{2}{\nano\second}$) broadband pulse as shown in
\cref{fig:signal_conditioning_raw}.
When digitizing this signal in its raw state, only very few samples can be
acquired. Further, especially the samples on the steep rising edge are highly
sensitive to slight variations in signal timing as may be introduced by for
example thermal influences on the transmission lines. This poses a problem for
long-term stability of the beam position measurements. This effect was observed
during a study using the KEK injector linac, the details of which are described
in \cref{sub:reseval_linac} (a).
Therefore, in order to maximize the number of samples acquired and mitigate
above mentioned effects, the signal is conditioned such as to stretch it as
much as possible. The upper limit for signal length is given by the
$\SI{96}{\nano\second}$ spacing of both bunches in two-bunch injection mode, as
the signal from first and second bunch may not overlap.

Stretching of the signal can be achieved by limiting its bandwidth by means of
a bandpass filter. The bandwidth is approximately inversely proportional to the
signal length and thus can be tuned by selecting a suitable bandpass filter.
While custom, optimized filters may be prepared eventually, for current
operation commercially available filters by Mini-Circuits are used.
A survey based on the S-parameter data published by Mini-Circuits showed that a
\textit{SBP-150+} bandpass or a combination of a \textit{SHP-200+} (highpass)
and \textit{SLP-200+} (lowpass) filters may provide suitable signal
conditioning. Initially four of each were purchased for testing.
When testing with actual signals during beam operation, the \textit{SBP-150+}
turned out to produce a shaped signal slightly longer than the
$\SI{96}{\nano\second}$ between injected bunches, leading us to adopt the
combination of \textit{SHP-200+} and \textit{SLP-200+}, which fulfilled the
spacing requirement.

Due to the filters, the signal is attenuated to a few tens of millivolts
peak-to-peak. Thus, to optimally utilize the input range of the ADCs, a MITEQ
\textit{AFS2-00100400-15-10P-4} $+\SI{20}{\decibel}$ amplifier is inserted into
the signal path of each channel, after the filters.
This amplifier was chosen as it not only fulfills or even exceeds the required
specifications but also was simply already available in sufficient quantity,
leftover from a previous project.

The shaped signal (i.e. with filters and amplifier inserted) for the case of
two-bunch injection is shown in \cref{fig:signal_conditioning_raw} (b). Note
that the timescale is different from the raw signal by more than an order of
magnitude. As the signal was recorded with the RFSoC 4x2 board, the y-axis
shows ADC counts rather than voltage.
Evidently the stretching of the signal is close to what is possible within the
constraint given by the $\SI{96}{\nano\second}$ bunch spacing.

\section{Firmware and Software}
\begin{figure}[htbp]
    \centering
    \includegraphics{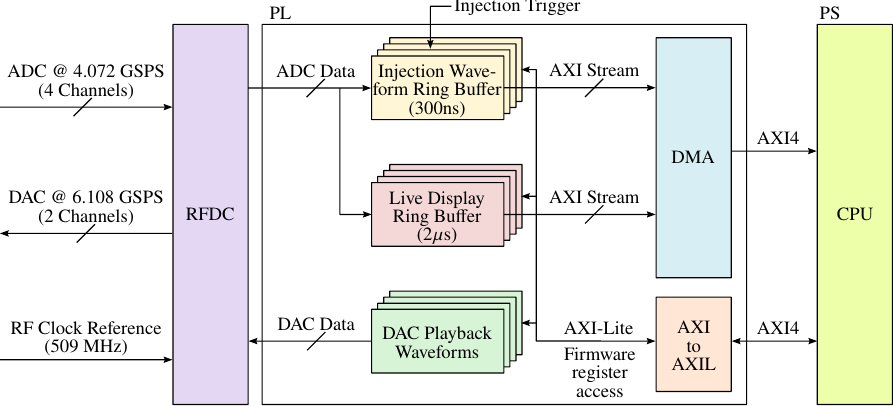} 
    \caption{Block diagram of the firmware run on the RFSoC.\label{fig:firmware_bd}}
\end{figure}

The firmware run on the RFSoC uses the \textit{axi-soc-ultra-plus-core}
framework developed at SLAC~\cite{slac:axi-soc-ultra-plus-core}.
A block diagram of the firmware is shown in \cref{fig:firmware_bd}.
The \textit{RF Data Converter} (RFDC) is the hardware block included in the
RFSoC chip containing the ADCs and DACs. It exposes RF inputs and outputs, a
sampling clock input as well as an interface to the FPGA, referred to in the
following as the Programmable Logic (PL).
We lock the RFDC sampling clock to the global RF clock of SuperKEKB at
$\SI{508.89}{\mega\hertz}$. The ADCs and DACs sampling frequency is an integer
multiple of this frequency. The sampling frequency is chosen as the highest
possible one satisfying this condition which is $\SI{4.065}{\giga\hertz}$.
While the DACs are not directly required for our application, we include
functionality to replay arbitrary waveforms in the firmware which is convenient
for testing purposes, e.g. loopback of DAC outputs into ADC inputs.

Data from the four available ADCs is continuously streamed to ring buffers
implemented in the PL. Ring buffers of two different sizes are available.
The larger one is sized at around $\SI{2}{\micro\second}$ and may be read out
for debugging and monitoring purposes at a slow rate (usually around
$\SI{1}{\hertz}$) if requested by software.
The smaller one is sized at around $\SI{300}{\nano\second}$ which is sufficient
for a recording of signals from both bunches in two-bunch injection mode,
including some margin. Readout of this buffer is triggered by the injection
trigger signal received from accelerator controls at maximally
$\SI{25}{\hertz}$ (although the firmware in principle may support much higher
trigger rates). A suitable delay is applied to the trigger in the firmware.

The CPU, here referred to as the Processing System (PS), runs an embedded Linux
distribution build using the Yocto Project tools.
Data from the ring buffers is streamed to the PS via Direct Memory Access
(DMA). For access to firmware registers a second interface is available mapping
the registers into the Linux memory space.
Stream DMA as well as register access use dedicated Linux kernel
drivers~\cite{slac:aes-stream-drivers}.

\begin{figure}[htbp]
    \centering
    \includegraphics{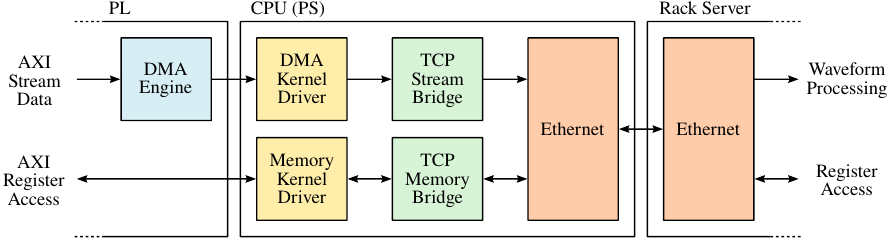} 
    \caption{Block diagram of the TCP bridge used for data transfer between the
    RFSoC board and a rack server running the waveform processing and device
    control code.\label{fig:software_bd}}
\end{figure}

As illustrated in \cref{fig:software_bd}, a TCP bridge application is run in
Linux userspace extending the stream and register interfaces over the local
network.
This utilizes the \textit{rogue} framework~\cite{slac:rogue} developed at SLAC,
providing an abstraction of the interfaces and high level access using Python
and/or C++.
Firmware control and waveform processing is handled by a rogue-based
application run on a rack server with all communication traversing the TCP
bridge. There is no restriction on the machine such an application may be run
on as long as it is connected to the same network as the RFSoC board, allowing 
for a flexible development flow.
In principle the application can also be run on the RFSoC's embedded CPU, which
however may lack the performance required for some more computationally heavy
tasks.
Some processing may also be implemented in PL, which however is not necessary
for our application with relatively low trigger rate.

The rack server application handles setup of the RFSoC board through the
register interface as well as processing of the incoming waveform data for each
injection.
The absolute value of the signal is integrated in a set time window for each
injected bunch. The obtained integral values are used as the signal strengths
$s_i$ and position is computed by minimizing \cref{eq:minimizefunc} as
described in \cref{sec:bpmchamber}.
While the control code is implemented in Python, the used minimizer is
implemented in C++ and invoked from Python code to ensure sufficiently fast
execution.

\section{Evaluation and Operation}
\subsection{Resolution estimation at the KEK injector linac}\label{sub:reseval_linac}
In order to verify basic operation and feasibility of the developed device, a
resolution study at the KEK injector linac was conducted during its operation
in June 2025.
This study used the stripline BPM chambers installed at the linac, which are,
as opposed to the SuperKEKB injection point BPMs, of much simpler geometry
(cylindrical with symmetric electrode arrangement).
Thus the pickup characteristics are expected to be quite different.
While the results of this study cannot be directly translated to the injection
point BPM chamber, they still are useful to judge whether performance of the
RFSoC based readout electronics is sufficient by comparing with the resolution
achieved by the readout electronics currently installed at the
linac~\cite{ichimiya:ibic2014-wepd04}.

During the study the importance of signal conditioning was not yet established
and no proper bandpass filters were available. However, to still somewhat
stretch the signal eventually \textit{Mini-Circuits SLP-450} lowpass filters
were added to the setup.

As only one RFSoC 4x2 board was available at the time, a resolution measurement
using the ordinary 3-BPM method~\cite{kekb_instr}, where three identical setups
are prepared and data is analyzed under the assumption that the resolution of
all setups is identical, was not possible.
Instead, data from two BPMs using the linac readout electronics was combined
with the data from a third one, located downstream from the other two, read out
using the RFSoC 4x2 board.
In this setup the assumption of equal resolutions is obviously not adequate.
Therefore the resolution for a BPM with linac readout electronics had to be 
determined separately using the ordinary 3-BPM method first, to then use it in
the resolution estimation for the RFSoC based readout. For the 3-BPM
measurement we used data from the same two BPMs as used for the RFSoC readout
resolution estimation, complemented by a third one located upstream of the
other two.

\begin{figure}[htbp]
    \centering
    \includegraphics[width=\textwidth]{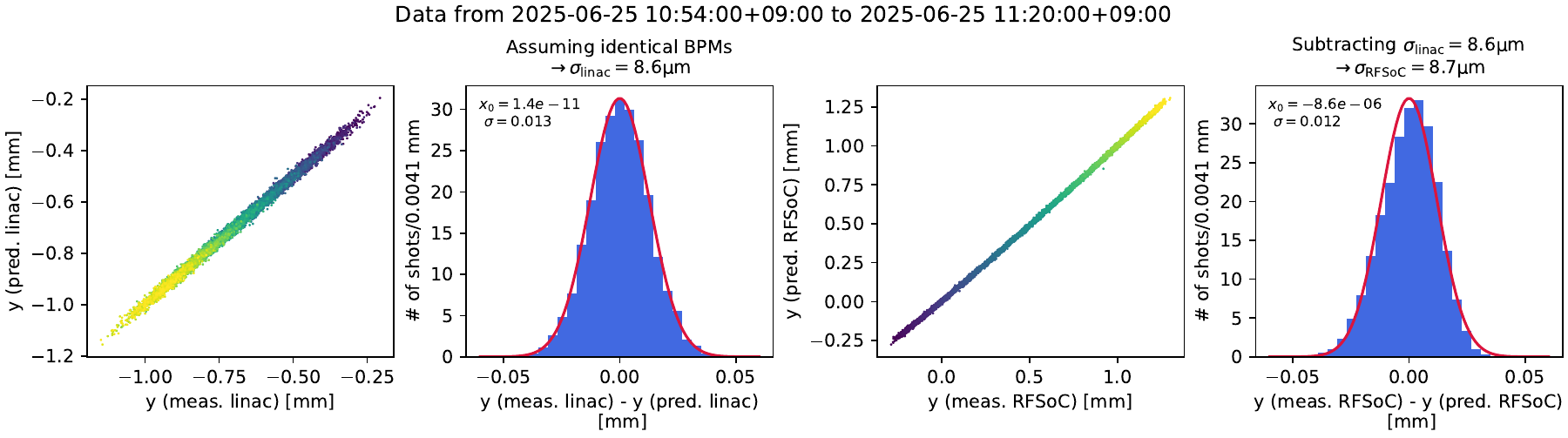}
    \caption{
        Predicted and measured vertical beam position used for resolution
        estimation. Due to the combination of RFSoC based and present linac
        readout electronics used, the resolution for the latter is determined
        first (left) and used in the resolution calculation of the former
        (right).
    \label{fig:linac_resolution_check}}
\end{figure}

To ensure a fair comparison, the data used for both the estimation of
resolution with linac BPM readout electronics as well as with the RFSoC based
readout electronics was taken from the same time window.
The beam position was scanned for around $\SI{1.5}{\milli\meter}$ in the
vertical ($y$) direction over the course of 25 minutes.
In both cases the resolution is determined from the width of the distribution
of differences between estimated and measured position at the third, most
downstream BPM.
The estimated position is determined by using the measured positions $y_1, y_2$
at the two upstream BPMs to linearly extrapolate the downstream beam position
as $y_3 = A y_1 + B y_2 + C$. The constants $A, B$ and $C$ are determined by a
fit over all available data.
For the fit to properly converge a range of positions is required. Thus only
the data for the vertical direction where the sweep was performed is used.
The estimated and measured positions as well as the distribution of their
differences are shown in \cref{fig:linac_resolution_check}.
The resolution is extracted from the standard deviation of the distribution of
differences.
For the case of only linac BPM readout electronics this is done under the
assumption of equal resolutions for all three setups
(\cref{fig:linac_resolution_check}, left) resulting in
$\sigma_{\text{linac}}=\SI{8.6}{\micro\meter}$.
For the measurement of the RFSoC based readout, this resolution is then
propagated through the linear extrapolation and eventually subtracted in
quadrature from the standard deviation of the corresponding distribution
(\cref{fig:linac_resolution_check}, right), resulting in $\sigma_{\text{rfsoc}}
= \SI{8.7}{\micro\meter}$.

The obtained resolution for the RFSoC based readout is within our expectations
of at least $\SI{10}{\micro\meter}$. Further it is comparable to the resolution
achieved by the linac readout electronics under the operating conditions during
the study.
It shall be noted that the linac BPMs paired with their readout electronics may
be capable of resolutions down to
$\SI{4}{\micro\meter}$~\cite{ichimiya:ibic2014-wepd04}, which however is known
to depend on operating conditions, thus rendering the observed resolution of
$\SI{8.6}{\micro\meter}$ still realistic.
Further, in terms of raw ADC performance, the RFSoC should outperform the linac
readout electronics but nevertheless did not achieve a resolution superior to
the latter. This however can be explained by the lack of proper signal
conditioning for the RFSoC based readout. The linac readout electronics contain
carefully tuned, purpose-built bandpass filters to condition the signal in
order to maximize the number of samples per pulse by stretching it as much as
possible within the limit of $\SI{96}{\nano\second}$ between bunches.
The lowpass filters used for the RFSoC based readout only stretched the signal
to around $\SI{20}{\nano\second}$ at best. If proper signal conditioning is
applied, around 4.5 times more samples can be acquired, which should improve
the resolution simply by statistical advantage.

A further artifact of the improper signal conditioning is the slight skew in
the distribution shown in \cref{fig:linac_resolution_check} (right). The short
signal and resulting lack of statistics combined with the rather steep edges of
the waveform make the position measurements sensitive to even very small
variations in relative phase of the signals from the electrodes (likely
influenced by temperature of the transmission lines, connectors etc.).
The result is a small drift of the position measurement becoming quite
noticeable when measuring for longer than around 30 minutes.
This effect was later reproduced on the bench, replaying signals of different
shapes using the DACs and confirmed to be negligible if proper signal
conditioning is applied.
However, it is clear that direct digitization of the raw broadband pulse is not
an option as for a BPM application, where long term stability of the measurement
is essential.

\subsection{Gain calibration}
\label{sec:gain_calibration}
A number of factors, including differences introduced by cables, filters etc.,
may influence the gain of each of the four channels.
As all those factors are difficult to evaluate separately, a beam data-driven
method is employed to correct for the resulting overall gain differences at
once.

We leverage on the fact that fit-based position computation method as described
in \cref{sec:bpmchamber} in principle also works with only three out of the
four available electrodes.
One may simply exclude the signal from one of the four electrodes from the
minimized function presented in \cref{eq:minimizefunc}. Excluding a signal is
in the following referred to as \textit{masking} it. There exist four masks
masking exactly one of the four channels.
If the overall gain for each channel is exactly the same, the four positions
computed with each of the four masks (in an ideal scenario) match exactly.
However, in the case of relative differences in channel gain, the four
positions will no longer match. 
Therefore, in the case of real (non idealized) data, the spread among the four
positions is expected to assume a minimum when the gain of each channel is
corrected such as to eliminate any relative gain differences between channels.

To quantify the spread among the four positions computed with the four masks
we use the quantity $r$, defined as described in the following.
Let $g_i$ where $i = 0,1,2,3$ be the correction factor applied to the signal from 
the $i$-th channel to obtain the corrected signal $s_i = g_i \cdot
s_i^{\text{raw}}$ used as an input to the position computation described in
\cref{sec:bpmchamber}.
We compute (as a function of the $g_i$) for each shot (index $j$) and direction
($d = x, y$) the variance $\sigma^2_{d, j}(\{g_i\})$ over the four values
obtained with the four masks. For $N$ shots this yields $2N$ values,
which are then summed to obtain the quantity $r$, which is to be minimized.
\begin{equation}
    r(\{g_i\}) = \frac{1}{2N}\sum_{d = x, y} \sum_{j = 0}^{N-1} \sigma^2_{d, j}(\{g_i\}).
\end{equation}
The $g_i$ minimizing this function are taken as the gain correction factors.

In practice, this minimization turned out to be non-trivial.
The dependence on the position computation, being itself based on a
minimization procedure using a rather complex signal strength map as an input,
makes the minimized quantity $r$ a considerably complex function of the gain
correction factors $g_i$. This significantly increases the chance for the
minimization to fail or get stuck in a local minimum. Such behavior was
observed for all attempted minimizers, even on idealized (generated) data. To
circumvent this, we first manually tune initial values for the $g_i$, judging
the spread of the positions computed with the masks by eye, by means of a
simple scatter plot. Using the such determined initial values, a minimizer is
usually able to find the correct global minimum. The choice of minimizer did
not matter much if initial values are chosen carefully. We settled on the
\textit{scipy}~\cite{2020SciPy-NMeth} implementation of the
\textit{Nelder-Mead} method. As the problem only has three degrees of freedom,
we fix $g_0=1$ and vary only the remaining three parameters.

Further complications arise when using real data recorded during accelerator
operation. First, during normal operations the injected beam usually traverses
the injection point BPM at around the same position, dictated by the trajectory
required for successful injection. Thus, the available data usually only covers
a small range (around a millimeter of spread) of positions. Studies using
idealized data showed that significant biases may be picked up when the covered
range of positions is this small. 

To avoid this, a dedicated study was conducted where the injection beam was
manually steered over an as large as possible range by manually adjusting the
septum magnets located directly upstream of the BPM. 
There is no beam shutter located downstream of the injection point BPMs,
so during the study, the injected beam inevitably enters the SuperKEKB storage
rings. 
As the trajectory of the injected beam is adjusted to far beyond what is used
during normal operations, the injected charge is unlikely to enter a stable
trajectory in the ring.
This in turn may lead to beam losses in the ring exceeding the thresholds set
for a beam abort and thus must be avoided. 
By conducting the study just after the start of the 2026 SuperKEKB operations,
where the rings were still operated with detuned optics and collimator
settings, providing a large acceptance for the injected beam, as well as
reducing the charge per injection and disabling one of the injection kickers in
the ring, losses could be suppressed sufficiently.

\begin{figure}[htbp]
    \centering
    \includegraphics[width=\textwidth]{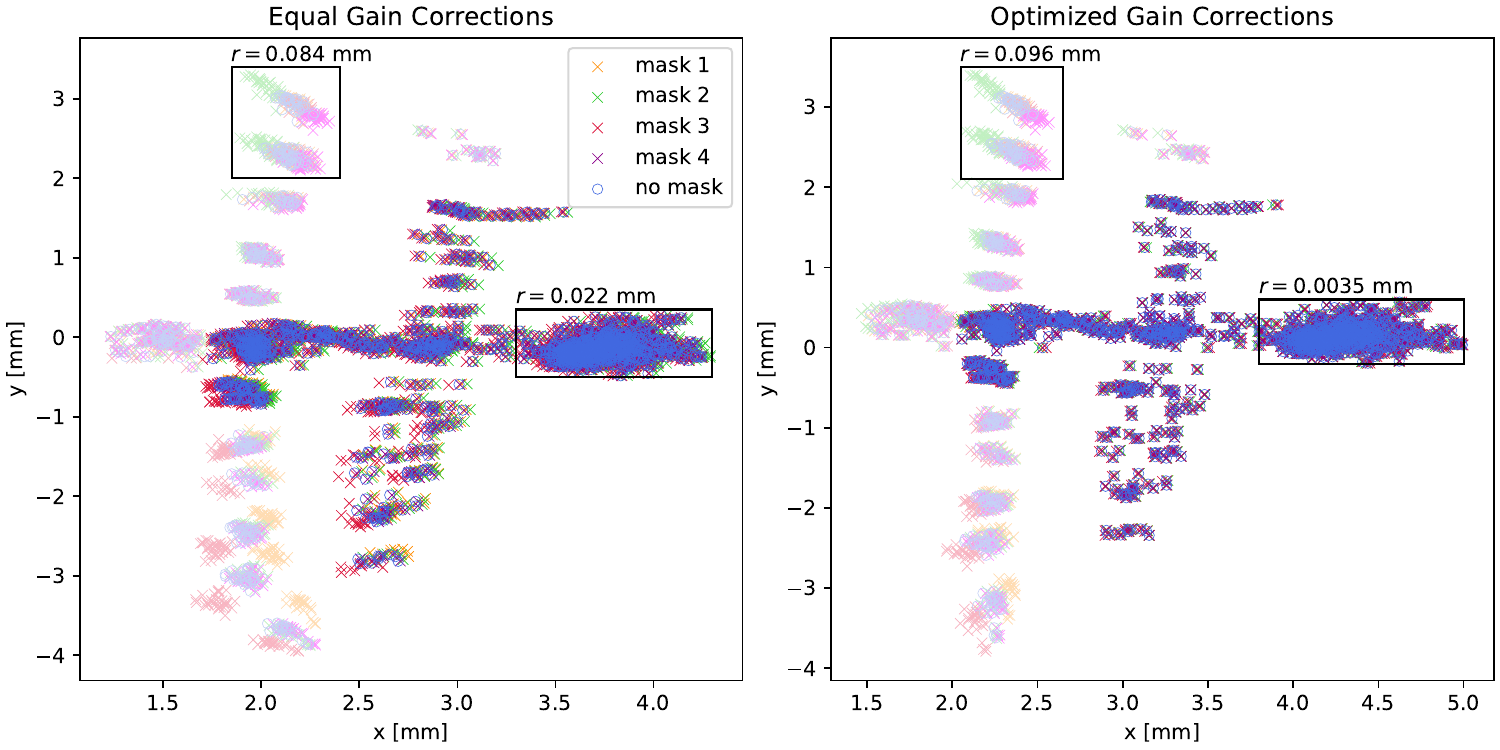}
    \caption{Beam positions recorded during the gain correction study, computed
        with each of the four possible masks as well as without any mask.
        Points excluded from the minimization of $r$ are shown in a lighter
        shade. Shown values of $r$ are computed over the regions indicated in
        black.
    \label{fig:gain_opt_plot}}
\end{figure}

The study was conducted for both of the electron and positron beams. 
Results for both beams are similar, although the region where satisfactory
accuracy could be achieved was much smaller for the positron beam. This is
understood to be related to the different electrode arrangement in the BPM
chamber. For beam positions very close to the electrodes, the four positions
computed with masks tend to spread out further, independent of the chosen set of
gain correction factors, indicating disagreement between simulation and true
signal response, as mentioned in \cref{sec:bpmchamber}.
For the positron beam, the electrode arrangement appears to amplify this effect.
As the case of the electron beam better illustrates our method and further the
chamber for the positron beam is scheduled to be replaced with a new one of the
same geometry as the one currently used for the electron beam
anyways\footnote{It is planned to perform a stretched wire scan of the new
chamber at this opportunity to better assess uncertainties of the currently
used BEM simulation.}, we restrict the following discussion to the case of the
electron beam.

The beam positions recorded during the study, computed with each mask as well
as without mask, for the case of equal gain correction factors (equivalent to
no gain correction applied) and optimized gain correction factors are shown in
\cref{fig:gain_opt_plot}.
For some regions, especially near the electrodes, it was found to be impossible
to reduce the spread of positions computed with masks much further than what is
shown for the case with optimized gain corrections. 
As this is understood to be a result of the discrepancy between BEM simulation
and true signal responses, the data from those regions is excluded from the
minimization process of $r$. The excluded points in \cref{fig:gain_opt_plot}
are plotted in lighter shades than those used for the minimization.

\Cref{fig:gain_opt_plot} further indicates the value of $r$ computed over two
regions, one near one of the electrodes and the other near the usual location
of the beam during operations, before and after the optimization. In the
former, $r$ does only slightly change (even increases) despite of a clear
reduction of $r$ in the latter. In this context, $r$ may be considered a
systematic uncertainty of the beam position. For the beam positions during
normal operations, $r\approx\SI{3.5}{\micro\meter}$, which is small compared
with the usual jitter of the beam.
Evidently gain compensation by the above presented method works well, as long
as the beam is not steered too close to one of the electrodes.

\subsection{Use during SuperKEKB operations}
The readout electronics for both rings of SuperKEKB were switched over to the
RFSoC based readout during the first weeks of the operational period starting
in November 2025.

The determined beam positions are provided to accelerator controls through the
\textit{Experimental Physics and Industrial Control System} (EPICS)
so called \textit{process variables} (PVs) published using
\textit{pythonSoftIOC}~\cite{softioc}, which was integrated into
the server side application. A much larger number of variables is
exposed through the rogue framework and can be accessed over the
local network.
\begin{figure}[htbp]
\centering
    \begin{subfigure}[t]{.48\textwidth}
        \centering\includegraphics[width=\linewidth]{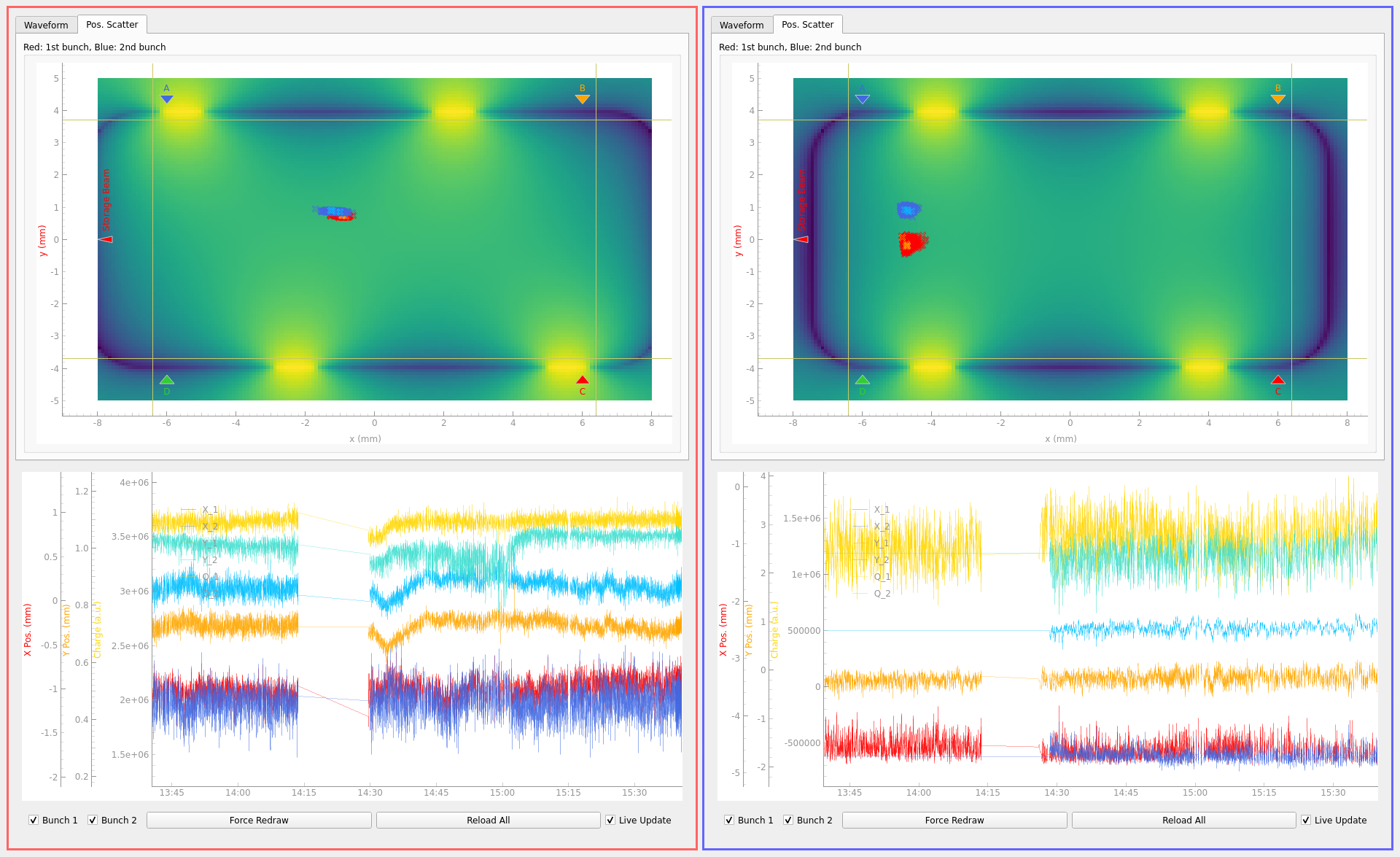}
        \caption{Position scatter plot view}
    \end{subfigure}
    \hfill
    \begin{subfigure}[t]{.48\textwidth}
        \centering\includegraphics[width=\linewidth]{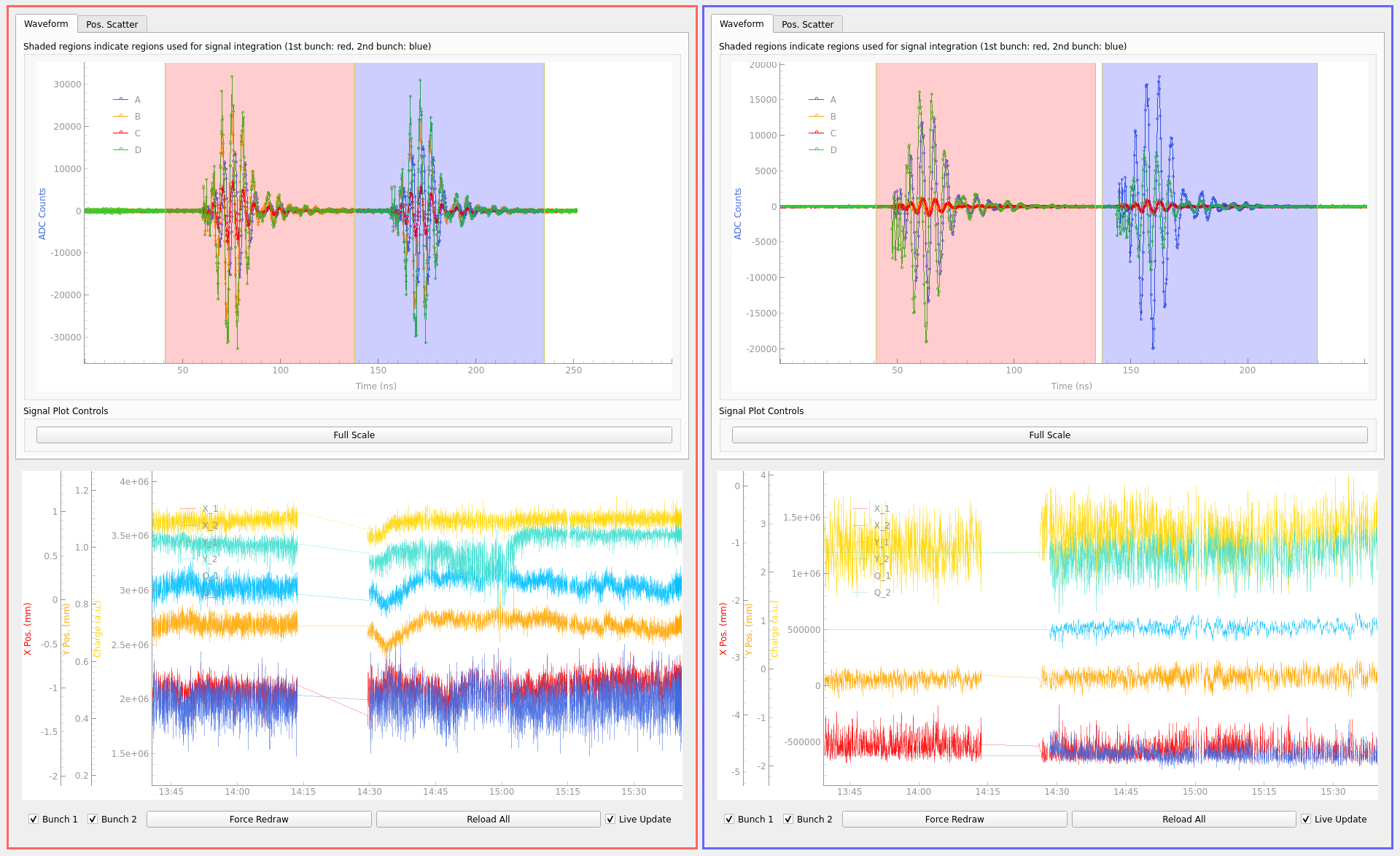}
        \caption{Waveform inspection view}
    \end{subfigure}
\caption{Screenshots of the GUI prepared for the control room.\label{fig:gui_screenshots}}
\end{figure}

A graphical user interface (GUI) or operator panel was prepared for use in the
SuperKEKB control room. This panel is based on \textit{pydm}~\cite{slac:pydm} and
fetches information from both EPICS PVs as well as rogue remote variables.
Screenshots of the panel are shown in \cref{fig:gui_screenshots}.
A \textit{scatter plot} tab, where the beam positions are plotted in near real
time, as well as a \textit{waveform} tab to inspect the raw waveforms is
available.
The former is plotted on top of the BEM map data summed over all four channels
to give some visual indication of the sensitivity of each electrode for a given
beam position as well as indicate the boundary of the beam duct.

\begin{figure}[htbp]
    \centering
    \includegraphics[width=0.5\textwidth]{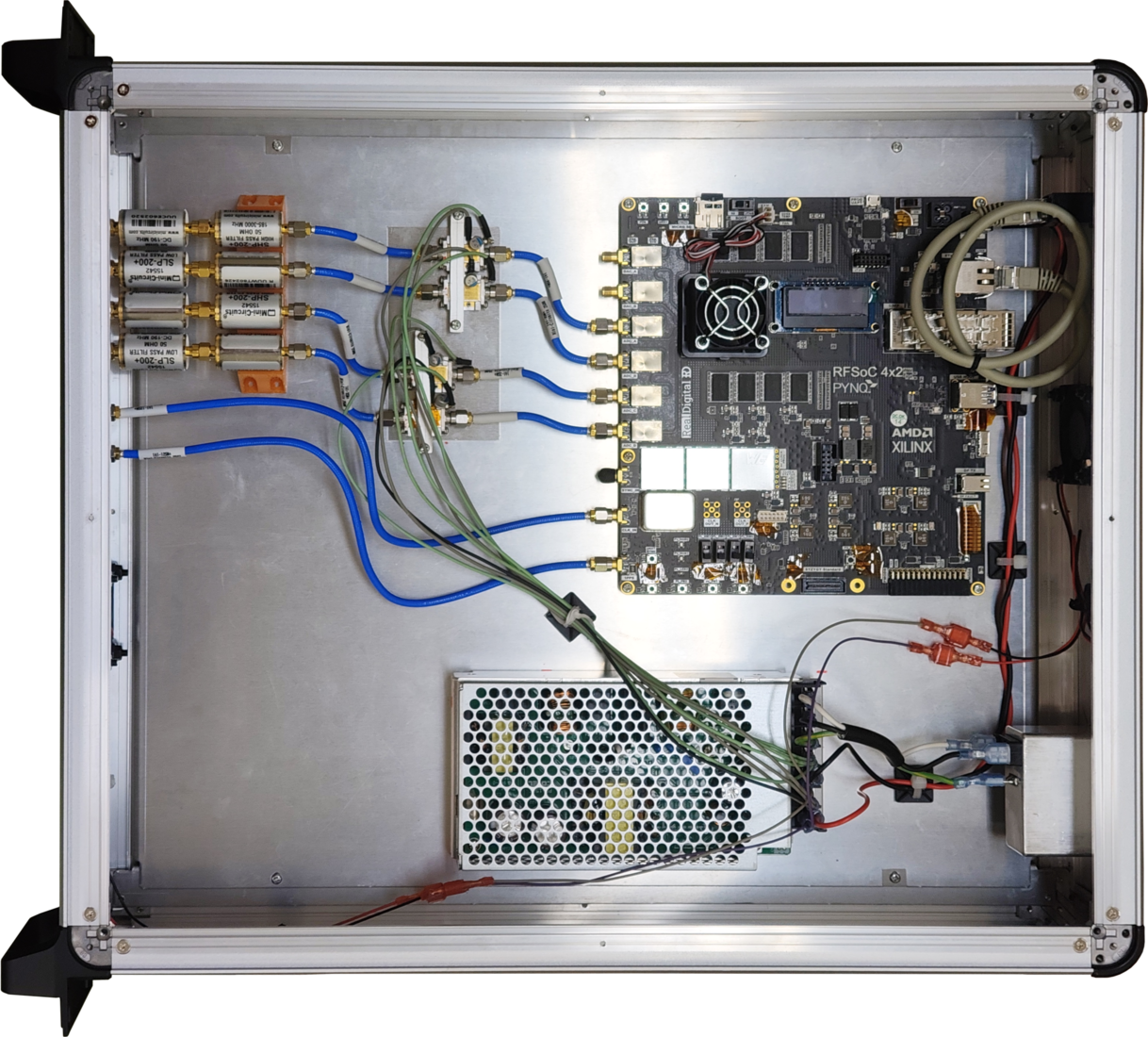}
    \caption{Photograph of the RFSoC 4x2 board, filters, amplifiers and power
        supply in the final configuration installed in a rack mounted
        enclosure.
    \label{fig:enclosure}}
\end{figure}

As no major complications have been encountered during the operations so far,
the initially temporary setup was moved to a proper rack mounted enclosure for
permanent installation during the 2025/2026 winter shutdown.
The final configuration is shown in \cref{fig:enclosure}

\section{Summary and Prospects}
In order to achieve its luminosity goals, SuperKEKB must sustain high beam
currents, which in turn requires continuous top-up injection.
To maximize the injected charge, tuning of the injection beam is required, an
important input to which is the position reading from the injection point BPMs.
We developed and successfully deployed a flexible solution for reliable
position measurement, specifically capable of independent measurement of both
bunches in the two-bunch injection mode, which with the previously used readout
device was only possible to a limited extent.

We adopted the RFSoC platform for this development with the secondary goal of
collecting experience with the technology towards future use for beam monitor
readout electronics at KEK.
The RFSoC platform in combination with the tooling and framework developed at
SLAC was found to provide a solid foundation enabling efficient and flexible
development of application specific, high performance readout solutions
involving RF sampling.

Studies towards possible adoption of RFSoC based readout at the \textit{beam
transport} line, connecting the injector linac to the SuperKEKB storage rings,
are already underway. As part of this, further features are being added to the
firmware, including integration with an event timing distribution system over
optical fiber as well as dynamic attenuator adjustment using the digital step
attenuators available in the 3rd generation RFSoCs.

Possible applications for use with fast bunch-by-bunch synchrotron radiation
monitors as well a large-scale deployment for a renewal of the readout
electronics of a subset of the BPMs installed in the SuperKEKB storage rings,
also involving a feedback system, are under consideration. 
For such applications, the currently used evaluation boards are no longer
suitable, which is why for future developments adoption of an RFSoC
\textit{System on Module} (SoM) centered platform is planned.

\section{Acknowledgments}
The work of B. Urbschat and G. Mitsuka was supported by Japan Society for the
Promotion of Science (JSPS) International Leading Research Grant Number
JP22K21347.
L. Ruckman's work was supported by the U.S. Department of Energy, under contract
number DE-AC02-76SF00515.

We would further like to thank F. Miyahara (KEK injector linac controls/beam
monitor group) and N. Iida (beam transport line group) for assisting with the
resolution study and collection of calibration data respectively.

\cleardoublepage


\bibliographystyle{JHEP}
\bibliography{biblio.bib}

\end{document}